# A LIGHTWEIGHT QR-ASSISTED ZERO-KNOWLEDGE IDENTIFICATION PROTOCOL FOR SECURE AUTHENTICATION

**Hüseyin BODUR**

Assistant Professor Dr., Düzce University, Faculty of Engineering, Department of Computer Engineering, Düzce,

ORCID ID: https://orcid.org/0000-0002-2815-3397

## ABSTRACT

This study proposes a lightweight Zero-Knowledge authentication model supported by QR codes. The approach is based on the Schnorr authentication protocol and provides an additional security layer against replay attacks through nonce and timestamp mechanisms. The proof data generated by the prover is embedded within a QR code and transmitted to the verifier. Thus, the system enables verification of knowledge of the secret key without revealing it. Simulation results show that proof generation and verification times under a 256-bit security level are in the millisecond range. Additionally, the proof size remains constant at approximately 0.5 KB, making it suitable for practical applications in terms of QR code capacity. The findings indicate that the proposed model is applicable in mobile and low-resource systems in terms of both security and performance.

**Keywords**: Zero-Knowledge Proof, Schnorr Protocol, QR Code Authentication

## ÖZET

Bu çalışmada, QR kod destekli hafif bir Zero-Knowledge kimlik doğrulama modeli önerilmektedir. Önerilen yaklaşım, Schnorr tabanlı kimlik doğrulama protokolünü temel almakta ve nonce ile zaman damgası mekanizmaları aracılığıyla yeniden oynatma (replay) saldırılarına karşı ek güvenlik katmanı sağlamaktadır. Kanıtlayıcı tarafta üretilen kanıt verisi QR kod içerisine gömülerek doğrulayıcıya iletilmektedir. Böylece sistem, gizli anahtarın ifşa edilmeden matematiksel olarak doğrulanmasına olanak tanımaktadır. Gerçekleştirilen simülasyon sonuçlarına göre, 256-bit güvenlik seviyesinde kanıt üretim ve doğrulama sürelerinin milisaniye seviyesinde olduğu gözlemlenmiştir. Ayrıca kanıt boyutunun yaklaşık 0.5 KB civarında sabit kaldığı ve QR kod kapasitesi açısından pratik uygulamalara uygun olduğu belirlenmiştir. Elde edilen bulgular, önerilen modelin hem güvenlik hem de performans açısından mobil ve düşük kaynaklı sistemlerde uygulanabilir olduğunu göstermektedir.

## INTRODUCTION

Secure authentication is crucial in today's mobile, distributed, and resource-constrained environments. Traditional authentication methods, such as password-based systems, one-time passwords, and certificate infrastructures, often require third-party systems to process sensitive user information. However, inadequate or poorly implemented authentication mechanisms introduce security risks, including identity theft, data leakage, and replay attacks [1-3].

Zero-Knowledge Proof (ZKP) protocols are robust cryptographic methods that enable a user to prove knowledge of confidential information without revealing it [4]. The Fiat-Shamir transformation is widely used in digital signature and authentication systems, as it converts interactive authentication protocols into non-interactive ones [5]. Schnorr-based authentication protocols, due to their mathematical security based on the difficulty of the discrete logarithm problem, are widely recognised in the literature for their security and computational efficiency in authentication and zero-knowledge proof applications [6-7]. However, their applicability in lightweight and mobile environments and their resilience to replay attacks in practical deployments remain limited.

Quick Response (QR) codes are widely used for rapid data transfer on mobile devices and in contactless authentication systems. QR-based systems are often preferred for event entry, payment, and check-in processes. However, most existing systems require a central server for verification, which creates serious security vulnerabilities if the QR code is copied or reused [8].

Numerous studies in the literature investigate ZKP-based or QR-code-based mechanisms for identity verification, data privacy, and secure access control. In one ZKP-based study, Moya et al. presented a privacy-protected authentication system called ZKBAR-V, which integrates zero-knowledge proofs with blockchain for academic record verification [9]. In another study, Maidine et al. proposed a blockchain-based identity management architecture for verifying IoT and identity claims, aiming to reduce dependence on centralised third parties by using ZKPs [10]. Al-Karawi and Akdeniz proposed a technique combining federated learning with ZKPs for privacy-protected authentication in 5G networks [11]. In another study, Patil et al. proposed an age verification system based on ZKP. This system enables identity verification without disclosing user data by using blockchain, a decentralised identity framework (SSI), verifiable credentials, and ZKPs [12].

Research on QR codes generally focuses on secure QR code generation and verification techniques to prevent tampering or counterfeiting. In his study, Alsuhibany proposed a system that protects QR code integrity through dynamic QR code generation and digital watermark integration [13]. In another study, Abbas analysed QR code-based two-factor authentication systems, aiming to improve the existing security of QR codes and reduce the risk of unauthorised access [14]. In their research, Sarkhi and Mishra developed a deep learning-based model for detecting and classifying cyberattacks using QR codes. They focused on identifying security threats by classifying whether the QR code content is malicious [15].

There are also studies in the literature where ZKP and QR code-based applications are used together. In one such study, Gokulakrishnan et al. proposed an identity and transaction verification system that utilises ZKP to ensure trust without disclosing confidential data. In this system, users and third parties can verify the validity of evidence using a QR code-based authentication mechanism that sends queries to the server endpoint without revealing private details [16]. In another study, Raipurkar et al. proposed a blockchain-based platform where real identities of users are stored in the relevant web applications using decentralised storage. On this platform, credentials are verified using ZKP and QR codes [17].

This study proposes a lightweight QR-enabled zero-knowledge authentication protocol that integrates QR-based data transmission with a Schnorr-based, non-interactive zero-knowledge authentication protocol. The model eliminates interactive verification steps through a Fiat-Shamir transformation and protects against replay attacks using a time-based verification mechanism. The generated zero-knowledge proof components are embedded within a dynamically generated QR code, resulting in a privacy-preserving verification process with low infrastructure requirements.

The main contributions of this study are as follows:

1. Design of a lightweight, non-interactive zero-knowledge authentication protocol that can be transmitted via QR code,
2. Reduction of replay risk through a timestamp-based verification mechanism,
3. Evaluation of the compatibility of the proof size with QR code capacity.

Unlike blockchain-based identity models, this study focuses on low computational cost, minimal infrastructure requirements, and suitability for mobile environments. It therefore offers a viable secure authentication solution for IoT gateway systems, temporary access control mechanisms, and mobile authentication applications. Experimental results show that the proposed protocol provides strong cryptographic security while maintaining a sufficiently low computational load for practical applications.

## PROPOSED PROTOCOL

### Parameters

p is a large prime and g is a generator of the multiplicative group modulo p. The prover selects a private key $x \in Z_{\{p-1\}}$ and computes the public key $y = g^x \bmod p$. The public key y is registered with the verifier before authentication.

### Prover Side (Proof Generation)

The prover refers to the party that possesses the private key (x) and wishes to verify its identity without revealing it. This component generates a random value (r) using a Schnorr-based zero-knowledge identification mechanism and calculates a temporary commitment value (t) accordingly. A challenge (c) is then created by inputting t, the nonce, and the timestamp into a hash function. Subsequently, the response value (s) is generated using r, c, and x. Throughout this process, x is never disclosed; only a mathematically verifiable proof is produced. Thus, the prover demonstrates identity without leaking information, relying on the computational hardness of the discrete logarithm problem. The generated proof is then encoded in QR format and prepared for transmission.

The prover performs the following operations in sequence:

1. Stores the private key x.
2. Generates a random value r.
3. Computes the temporary commitment value $t = g^r \; mod \; p$ .
4. Calculates the challenge value $c = Hash(t \;||\; y \;||\; nonce \;||\; timestamp)$.
5. Calculates the response $s = r + cx \; mod \; (p - 1)$.
6. Embeds the proof data $\pi = (t, s, nonce, timestamp)$ in JSON format into a QR code.

**Verifier Side**

The verifier is the validation component that mathematically checks the accuracy of the proof generated by the prover. It decodes the proof transmitted via QR code, checks the validity of the timestamp, and reconstructs the value of c using a hash function. It then tests the Schnorr validation equation $g^s = t \cdot y^c \; mod \; p$. If the equation holds, the proof is accepted; otherwise, it is rejected. As the verifier performs validation based solely on the public key, the system maintains zero-knowledge functionality. In addition, time tolerance control protects against replay attacks.

The verifier performs the following operations in sequence:

1. Reads the proof (π) from the QR code.
2. Performs a timestamp check.
3. Calculates the challenge value $c = Hash(t \;||\; y \;||\; nonce \;||\; timestamp)$ .
4. Checks the equality: $g^s = t \cdot y^c \; mod \; p$ .
5. Gives an Accept or Reject decision.

**QR channel**

A QR channel is a carrier layer that enables the transmission of generated zero-knowledge proofs over a physical or optical medium. This channel allows proof data encoded in JSON format to be transferred between two parties without a network connection. The QR mechanism does not provide cryptographic security; it only enables one-way data transmission while preserving the integrity of the proof. However, as QR codes are easily copied, there is a risk of replay, so the system implements a timestamp and nonce mechanism for freshness verification. This approach allows the QR channel to offer a practical and contactless transmission solution without compromising security.

## SECURITY & PERFORMANCE ANALYSIS

### Security Analysis

The proposed QR-enabled zero-knowledge authentication protocol is based on a Schnorr identification mechanism and derives its security from the computational difficulty of the Discrete Logarithm Problem (DLP). The protocol's key security features are examined below.

*Completeness*

If the prover knows the secret key x and follows the protocol steps correctly, the verifier will satisfy the equation $g^s = t \cdot y^c \; mod \; p$. This equation is derived from $s = r + cx \; mod \; (p - 1)$. Therefore, verification always succeeds between honest parties.

*Soundness*

If the attacker does not know the secret key x, satisfying the validation equation requires solving a discrete logarithm problem. The probability that an adversary can satisfy the verification equation without knowledge of the secret key is negligible under the hardness assumption of the discrete logarithm problem. As the challenge value generated by the hash function is deterministic and unpredictable, it is practically impossible for the attacker to produce a valid proof in advance.

*Zero-Knowledge Property*

The protocol does not reveal any information about the secret key. Extracting the secret key from the generated proof values requires solving a discrete logarithm problem. Furthermore, as the challenge value is generated using a hash function, a non-interactive version of the interactive structure (the Fiat–Shamir transformation) is applied.

*Replay Attack Resistance*

Due to the visually copyable nature of QR codes, replay attacks present a significant threat. The proposed model addresses this risk through two mechanisms:

A timestamp, indicating when the proof was generated, enables verification of its validity window during the authentication process. The verifier compares the timestamp in the proof with the current system time to determine if it falls within a specified tolerance range (Δ). If the time difference exceeds a predefined threshold, the proof is rejected. This method adds an extra layer of protection against delayed replay attacks. The timestamp offers temporal context, ensuring the proof is valid only within a specific time frame.

A nonce is a randomly generated, single-use value for each authentication session. Its primary purpose is to ensure that proofs generated with the same secret key and similar timing remain distinct (proof diversity). Proofs with manipulated timestamps are rejected during verification.

**Performance Analysis**

The protocol's performance was analysed using a Python-based simulation environment. The implementation utilises modular exponentiation operations of the Schnorr protocol and the SHA-256 cryptographic hash function for challenge generation. The analysis evaluates the computational cost of the cryptographic operations and the size of the generated proof.

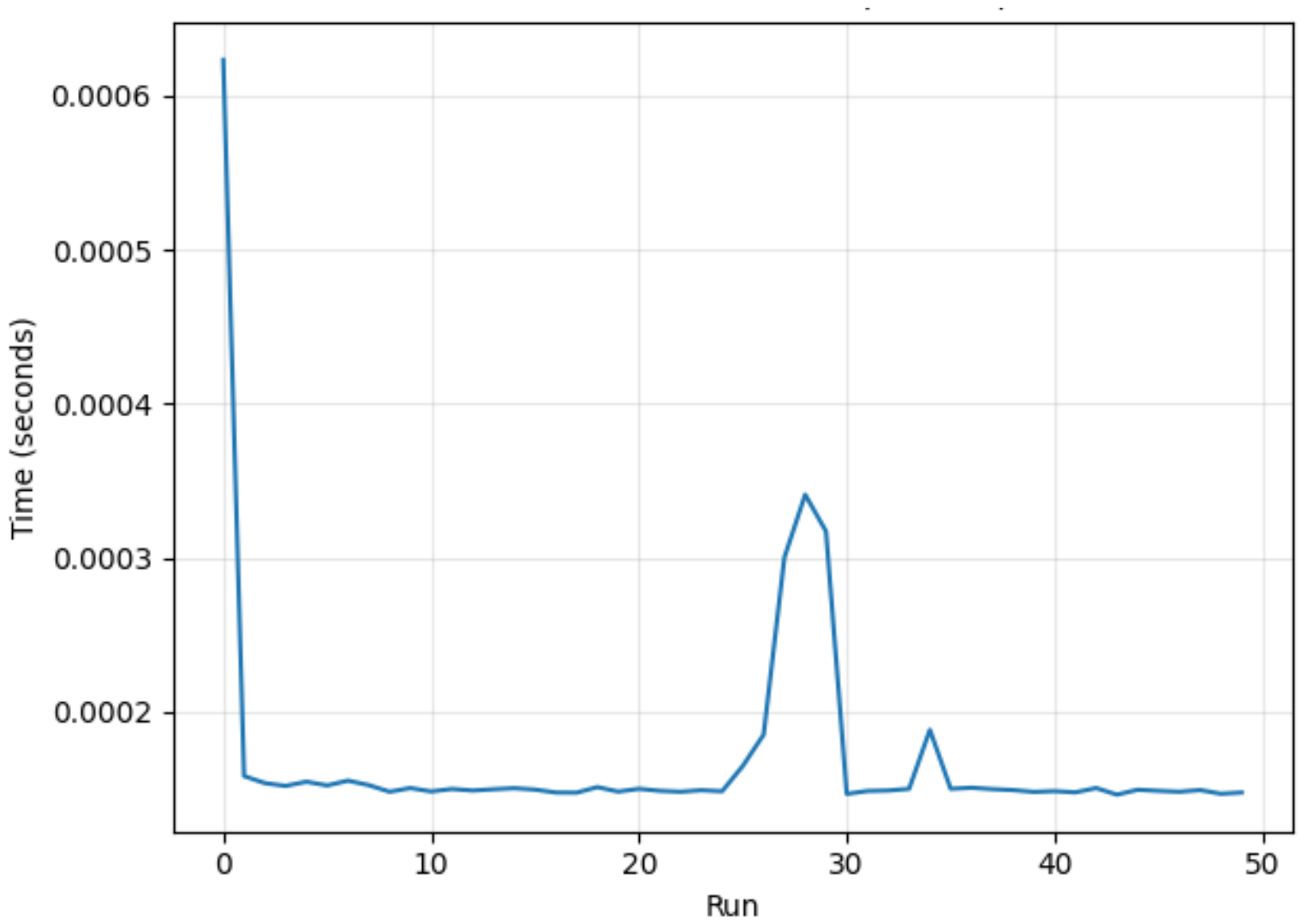

Figure 1.Proof Generation Time (256-bit)

Figure 1 shows the proof generation time on the prover side over 50 iterations. The results indicate that the generation time generally ranges from 0.00014 to 0.00018 seconds. The higher time observed in the first run is due to the costs of prime number generation and parameter initialisation. Similarly, small increases observed during some iterations may result from the variability of random number generation and modular exponentiation operations, depending on system load. Overall, the generation time is lower than or similar to the verification time. This suggests that the system is computationally lightweight on the client side and suitable for real-time authentication scenarios.

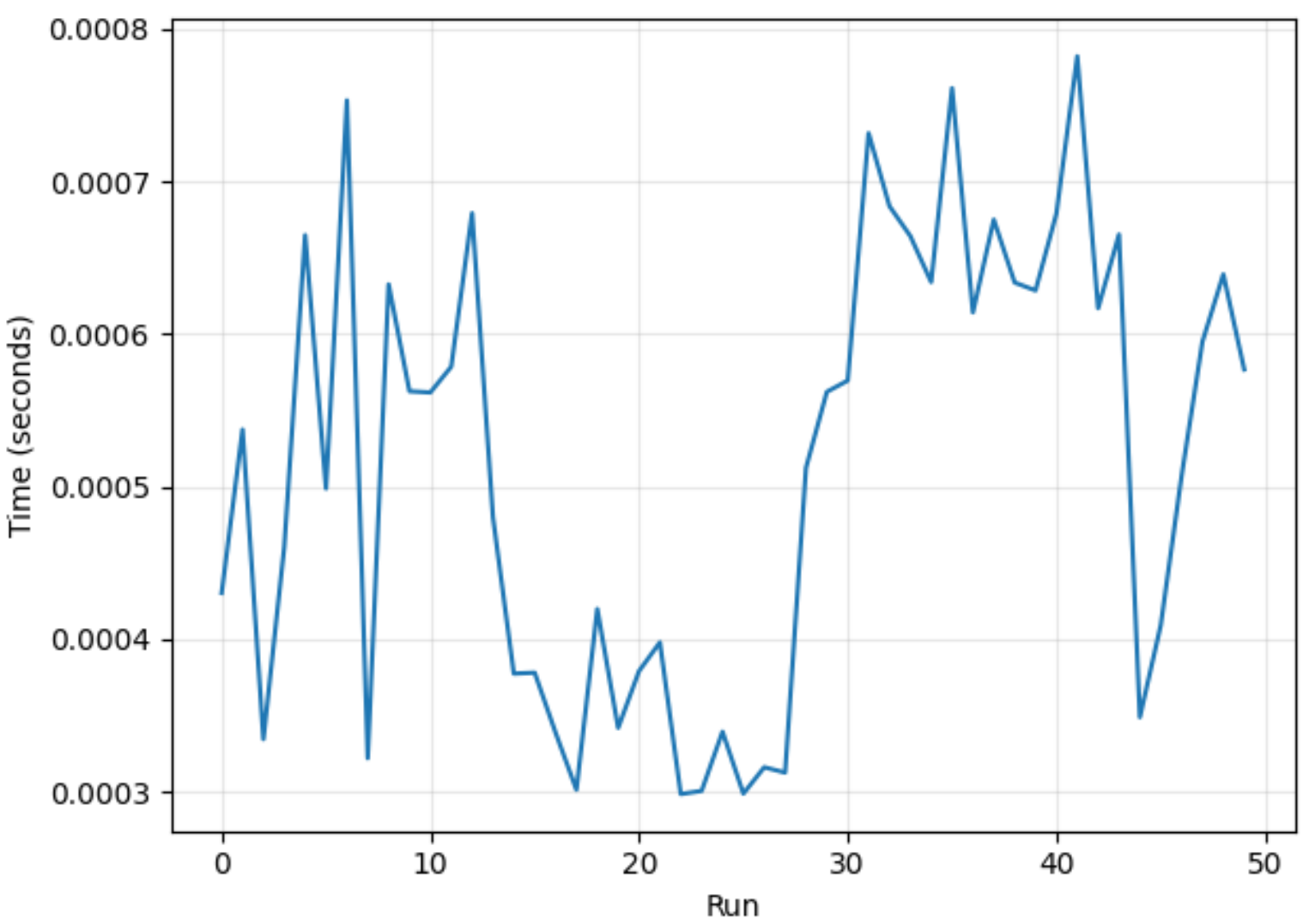

Figure 2.Proof Verification Time (256-bit)

Figure 2 shows the validation time of a zero-knowledge proof under a 256-bit security parameter over 50 independent runs. The horizontal axis indicates the number of test iterations, and the vertical axis shows the validation time (in seconds). The results indicate that the validation time ranges approximately from 0.00030 to 0.00078 seconds. Minor fluctuations in execution time are mainly attributable to system load variations, randomness in parameter generation, and Python interpreter overhead. Overall, the validation time remains below one millisecond. This demonstrates that the Schnorr-based validation process is computationally lightweight, requiring only two modular exponentiation steps and one hash calculation. This result is particularly significant for applicability to mobile or low-end verifier devices.

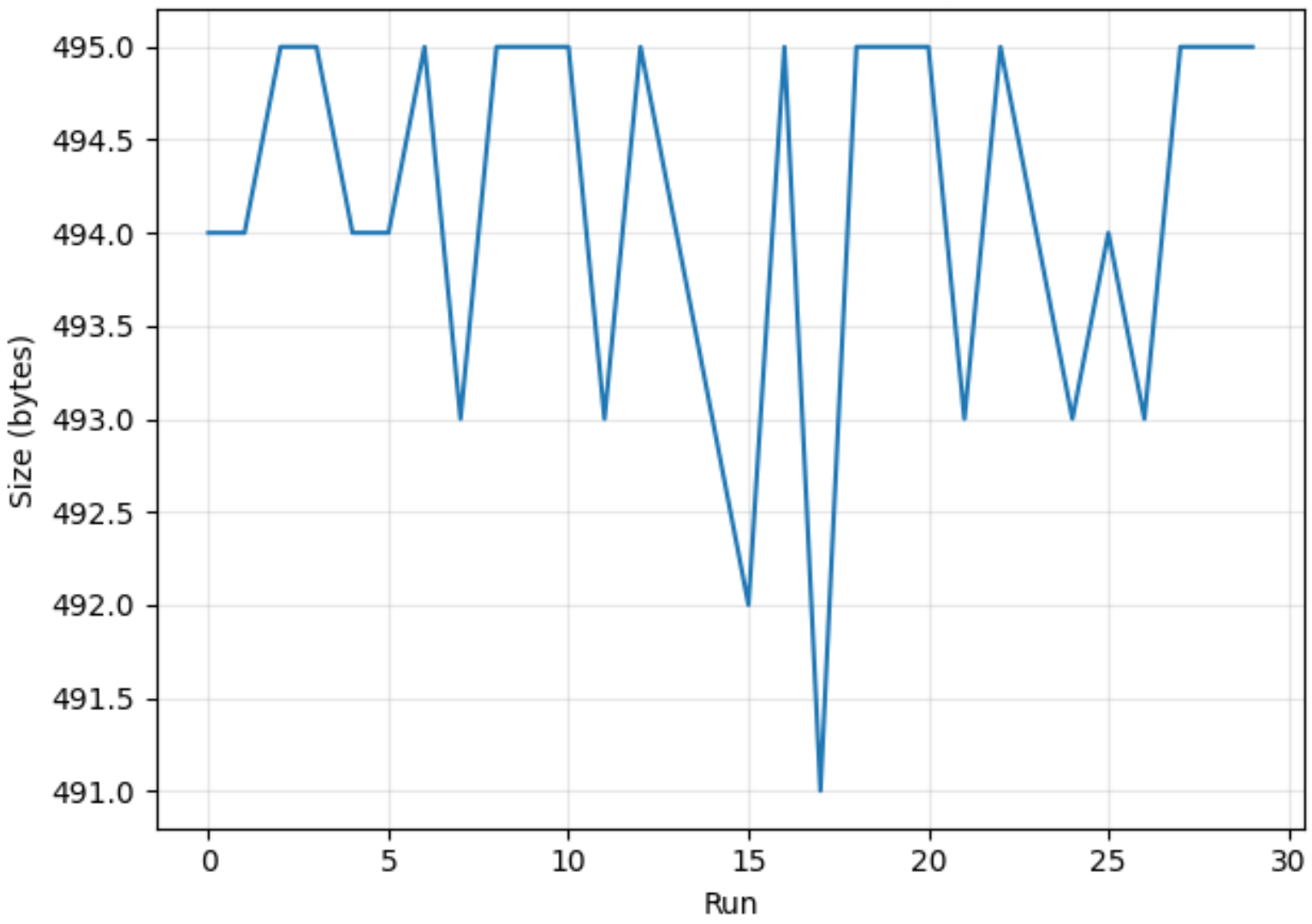

Figure 3.Zero-Knowledge Proof Size (bytes)

Figure 3 shows the size in bytes of the generated proof data after encoding in JSON format. The horizontal axis shows the different generation attempts, while the vertical axis indicates the proof size (bytes). The proof size remains constant at approximately 491–495 bytes. This result suggests that the proof structure comprises deterministic-length cryptographic parameters and that the data size does not vary significantly from sessions. This data, at approximately 0.5 KB, is well suited to QR code capacity and can be easily transmitted even with lower-version QR codes. This supports the system's applicability in practical deployment scenarios, such as physical access control and mobile authentication.

## CONCLUSION

In this study, a lightweight, QR-supported Zero-Knowledge authentication protocol is designed and evaluated for performance. The proposed model employs a Schnorr mechanism based on the difficulty of the discrete logarithm problem to enable verification without revealing the secret key. Session freshness is ensured through nonce and timestamp components, providing additional resilience against replay attacks.

Simulation results indicate that the system is suitable for practical applications, offering millisecond-level processing times and a fixed proof size. The proposed model is considered a viable solution, particularly for QR code-based physical access systems, mobile authentication scenarios, and devices with low computational capacity.

This study introduces a lightweight and practical authentication architecture suitable for QR-based verification environments. Future work will focus on integrating the model with blockchain-based identity management infrastructures, conducting scalability analyses, and performing experimental tests on real hardware. In this way, the role of the proposed approach in distributed authentication systems can be evaluated more comprehensively.